\documentclass[preprint,12pt]{elsarticle}

\usepackage{amssymb}
\usepackage{amsmath}
\usepackage{amsthm}
\usepackage{amsfonts}
\usepackage{mathrsfs}
\usepackage{t1enc}
\usepackage{graphicx,color}
\usepackage[usenames,dvipsnames,svgnames,table]{xcolor}
\usepackage{epstopdf}
\usepackage{caption}
\usepackage{subfig}

\newtheorem{remark}{Remark}

\newcommand{\NN}{\mathbb{N}}
\newcommand{\ZZ}{\mathbb{Z}}

\journal{Physica D}

\definecolor{DarkRed}{rgb}{0.45,0.1,0.1}

\begin{document}

\begin{frontmatter}



\title{Controlling roughening processes in the stochastic Kuramoto-Sivashinsky equation}


\author{S. N. Gomes$^1$,  S. Kalliadasis$^2$, D. T. Papageorgiou$^1$, G. A. Pavliotis$^1$, and M. Pradas$^3$}
\address{$^1$Department of Mathematics, Imperial
College London, London, SW7 2AZ, UK\\
$^2$Department of Chemical Engineering, Imperial
College London, London, SW7 2AZ, UK\\
$^3$School of Mathematics and Statistics, The Open University, Milton Keynes MK7 6AA, UK
}

\begin{abstract}
We present a novel control methodology to control the roughening processes
of semilinear parabolic stochastic partial differential equations in one dimension,
which we  exemplify with the stochastic Kuramoto-Sivashinsky equation.
The original equation is split into a linear stochastic and a nonlinear deterministic
equation so that we can apply linear feedback control methods. Our control strategy
is then based on two steps: first, stabilize the zero solution of the deterministic
part and, second, control the roughness  of the stochastic linear equation. We
consider both periodic controls and point actuated ones, observing in all
cases that the second moment of the solution evolves in time according to a
power-law until it saturates at the desired controlled value.
\end{abstract}

\end{frontmatter}

\section{Introduction}
\label{Sec:Intro} Roughening processes arise in nonequilibrium systems due
to the presence of different mechanisms acting on multiple time and
lengthscales and are typically characterized by a time-fluctuating ``rough"
interface whose dynamics are described in terms of a stochastic partial
differential equation (SPDE). Examples are found in a broad range of
different applications, including surface growth dynamics such as
e.g.~surface erosion by ion sputtering
processes~\cite{Cuerno1995,Cuerno1995a}, film deposition in electrochemistry~\cite{Buceta1997,Buceta1998}, or by other methods~\cite{Hu2008a,Hu2008b},fluid flow in porous
media~\cite{Alava2004,Soriano2005,Pradas2006},  fracture
dynamics~\cite{Bouchbinder2006} and thin film
dynamics~\cite{Kalliadasis2012,Diez2016,Nesic2015,Gruen2006,Bloemker2002}, to name but
a few. Not surprisingly, understanding the dynamics of the fluctuating
interface in terms of its roughening properties, which often exhibit
scale-invariant universal features and long-range spatiotemporal
correlations, has become an important problem in statistical physics which
has received considerable attention over the last
decades~\cite{Barabasi1995}. In addition, the ability of controlling not only
the dynamics of the  surface roughness (e.g.~ its growth rate) but also its
convergence towards a desired saturated value has recently received an increased
interest due to its applicability in a wide spectrum of natural phenomena and
technological applications.

Here we  present a generic linear control methodology for controlling the
surface roughness, i.e., the variance of the solution, of nonlinear SPDEs
which we  exemplify with the stochastic Kuramoto-Sivashinsky (sKS) equation.
The starting point is to split the original SPDE into a stochastic linear part and a
deterministic nonlinear part, and to apply existing control
methodologies~\cite{IMA_paper,PRE_paper} to the nonlinear deterministic part.
Our control strategy is based on two steps: first, stabilize the
zero solution of the deterministic system and, second, control the second
moment of the solution of the stochastic linear equation (e.g.~a measure of
the surface roughness)  to evolve towards any desired value. By considering
either periodic or point actuated controls, our results show that the second
moment of the solution grows in time according to a power-law with a
well-defined growth exponent until it saturates to the prescribed value we
wish to achieve.

It is important to note that other control strategies have been proposed
previously for controlling the surface roughness and other quantities of
interest, such as the film porosity and film thickness in various linear
dissipative models, including the stochastic heat equation, the linear sKS
equation, and the Edwards-Wilkinson (EW) equation; see e.g.~
\cite{Hu2008a,Hu2008b,Hu2009,Hu2009a,Lou2005,Lou2006,Lou2008,Lou2009,Zhang2010}.
However, it should also be emphasized that  most of these works involve the
use of nonlinear feedback controls which change the dynamics of the system
and require knowledge of the  nonlinearity at all times, something that may
be difficult to achieve. We believe that our framework offers several
distinct advantages since the controls we derive and use are linear functions
of the solution which do not affect the overall dynamics of the system and
also decrease the computational cost. Another recent study is
Ref.~~\cite{Harrison2016} which considered a deterministic version of the KS
equation, and presented a numerical study of the effects of the use of ion
bombardment which varies periodically in time on the patterns induced by the
ion beams on an amorphous material. In particular, this study found that
rocking the material sample about an axis orthogonal to the surface normal
and the incident ion beam, which corresponds to making the coefficients of
the KS equation periodic in time, can lead to suppression of spatiotemporal
chaos.

The work presented in this paper is motivated by earlier research carried out by our group: on one hand,
the study of noise induced stabilization for the Kuramoto-Shivashinsky (KS) equation~\cite{Pradas2011, Pradas2012} and, on the other hand, the study of optimal and feedback control methodologies for the KS equation and related
equations that are used in the modeling of falling liquid films~\cite{IMA_paper,PRE_paper,Thompson2016}. It was shown
in~\cite{Pradas2011, Pradas2012} that appropriately chosen noise can be used in order to suppress linear instabilities in
the KS equation, close to the instability threshold. Furthermore, it was shown in~\cite{IMA_paper,PRE_paper} that nontrivial
steady states and unstable traveling wave solutions of the deterministic KS equation can be stabilized using appropriate
optimal and feedback control methodologies. In addition, similar feedback control methodologies can be used in order to
stabilize unstable solutions of related PDEs used in the modeling of falling liquid films, such the Benney and the weighted-residuals equations.

The rest of the paper is structured as follows.
Section 2 introduces the sKS equation and discusses means to characterize the
roughening process of its solution. In Section 3 we outline the general
linear control methodology which is applied to the case of periodic controls
in Section 4, and point actuated controls in Section 5.
A summary and conclusions are included in Section 6.

\section{The stochastic Kuramoto-Sivashinsky (sKS) equation}
Consider the sKS equation:
\begin{equation}\label{sKS}
u_t = -\nu u_{xxxx} - u_{xx} - uu_x + \sigma\xi(x,t),
\end{equation}
normalized to $2\pi$ domains ($x \in [0,2\pi]$) with $\nu=(2\pi/L)^2>0$,
where $L$ is the size of the system, with periodic boundary conditions (PBCs)
and initial condition $u(x,0) = \phi(x)$. $\xi(x,t)$ denotes Gaussian mean-zero spatiotemporal noise, which is taken to be white in time, and whose strength is controlled by the
parameter $\sigma$:
\begin{equation}\label{covariance}
\left<\xi(x,t)\xi(x',t')\right> = \mathcal{G}(x-x')\delta(t-t'),
\end{equation}
where $\mathcal{G}(x-x')$ represents its spatial correlation function. We can, in principle, consider the control problem for SPDEs of the form~\eqref{sKS} driven by noise that is colored in both space and time. Such a noise can be described using a linear stochastic PDE (Ornstein-Uhlenbeck process)~\cite{Prato2014}.

The noise term can be expressed in terms of its Fourier components as:
\begin{equation}\label{eq:noise_Fourier}
\xi(x,t) = \sum_{k=-\infty}^{\infty} q_k\,\dot{W}_k(t)\,e^{ikx},
\end{equation}
where $\dot{W}_k(t)$ is a Gaussian white noise in time and the coefficients
$q_k$ are the eigenfunctions of the covariance operator of the noise. For
example, if $\mathcal{G}(x-x') = \delta(x-x')$ (which corresponds to
space-time white noise), we have $q_k=1$. For the noise to be real-valued, we
require that the coefficients $q_k$ verify $q_{-k} = q_k$. Proofs of existence and uniqueness
of solutions to Eq.~(\ref{sKS})  can be found in
\cite{Duan2001,Ferrario2008}, for example. The behavior of Eq.~(\ref{sKS}) as a
function of the noise strength, and for particular choices of the coefficients $\{q_k\}$ has been analyzed in detail
in~\cite{Pradas2011,Pradas2012}. In particular, it was shown that sKS
solutions undergo several state transitions as the noise strength increases, including critical on-off
intermittency and stabilized states.

The quadratic nonlinearity in Eq.~(\ref{sKS}) is typically referred to as a Burgers nonlinearity.
We note that an alternative version of  Eq.~(\ref{sKS}) is found by making the change of variable $u = -h_x$,  giving rise to
\begin{equation}\label{sKS KPZ}
h_t = -\nu h_{xxxx} - h_{xx} +\frac12 (h_x)^2 + \sigma\eta(x,t),
\end{equation}
where $\xi(x,t) = \partial_x \eta(x,t)$. The main effect of this
transformation is to change the dynamics of the average
$u_0(t)=\frac{1}{2\pi}\int_0^{2\pi}u(x,t)\,dx$ of the solution. Indeed, Eq.~(\ref{sKS}) with
PBCs  preserves the value of $u_0$
whereas as a consequence of the nonlinear term $(h_x)^2$, Eq.~(\ref{sKS KPZ}) does not
conserve the mass $h_0(t)=\frac{1}{2\pi}\int_0^{2\pi}h(x,t)\,dx$.
Both equations have received a lot attention over the last decades, with
Eq.~(\ref{sKS}) more appropriate in mass-conserved systems such as the dynamics of thin
liquid
films~\cite{Pradas2012,Kalliadasis2012,Diez2016,Nesic2015,Gruen2006,Bloemker2002}, and
Eq.~(\ref{sKS KPZ}) relevant in modeling surface growth processes such as surface
erosion by ion sputtering
processes~\cite{Buceta1997,Buceta1998,Cuerno1995,Cuerno1995a,
Lauritsen1996,Lou2006,Rost1995}. It is also worth mentioning that the quadratic
nonlinearity appearing in Eq.~(\ref{sKS KPZ}) is the same as that in the
Kardar-Parisi-Zhang (KPZ) equation~\cite{Kardar1986,Hairer2013}
\begin{equation}\label{KPZ}
h_t =  h_{xx} +\frac12 (h_x)^2 + \sigma\eta(x,t).
\end{equation}
In fact extensive work indicates that Eq.~(\ref{sKS KPZ})
and Eq.~(\ref{KPZ}) are asymptotically equivalent, something referred to as
the ``Yakhot conjecture"~\cite{Yakhot1981,Procaccia1992,Elezgaray1994}.
Throughout the remainder of this study we will refer to Eq.~(\ref{sKS}) as
the sKS equation with Burgers nonlinearity and Eq.~(\ref{sKS KPZ}) as the sKS
equation with KPZ nonlinearity.

\subsection{Surface roughening}

An important feature of systems involving dynamics of rough surfaces is that
one often observes the emergence of scale invariance both in time and space,
i.e., the statistical properties of quantities of interest are described in
terms of algebraic functions of the form $f(t)\sim t^\beta$ or $g(x)\sim
x^\alpha$, where $\alpha$ and $\beta$ are referred to as scaling
exponents. An example of this is the surface roughness, or variance of
$u(x,t)$, which is defined as
\begin{equation}\label{Roughness}
r(t)=\sqrt{\frac{1}{2\pi}\int_0^{2\pi}\left[u(x,t) - u_0(t)\right]^2 \ dx}.
\end{equation}
We remark that $u_0$ may or may not depend on time, depending on whether
we consider the Burgers or the KPZ nonlinearities. Usually the above
quantity grows in time until it reaches a saturated regime, in which the
fluctuations become statistically independent of time and are scale-invariant
up to some typical length scale of the system, say $\ell_s$. This behavior
can be expressed as:
\begin{equation}
\label{Beta}
\langle r(t)\rangle \sim  \left\{\begin{array}{ll}
          t^\beta &  \mathrm{if}\ \ t\ll t_s, \\
          r_s &   \mathrm{if}\ \ t\gg t_s, \end{array}\right.
\end{equation}
where $\langle \dots\rangle$ denotes average over different realizations,
$\beta$ is the so-called growth exponent~\cite{Barabasi1995}, and $t_s$ and
$r_s$ are the saturation time and saturated roughness value, respectively,
which  depend on the length scale $\ell_s$. In particular, at a given time
$t<t_s$, the correlation of these fluctuations are on a spatial length scale
which grows in time as $\ell_c\sim t^{1/z}$. Therefore, saturation occurs
whenever $\ell_c = \ell_s$ from which we find $r_s\sim \ell_s^\alpha$ with
$\alpha=\beta z$. In this context, the exponents $\alpha$ and $z$ are the
roughness and dynamic exponent, respectively, and their particular values
determine the type of universality class~\cite{Krug1997}. For example, it is
known that the long-time behavior of the KPZ equation Eq.~(\ref{KPZ}), is
characterized by the  KPZ universality class with  $\alpha = 1/2$ and
$z=3/2$, while its linear version, which is referred to as the
Edwards-Wilkinson (EW) equation, is characterized by the EW universality
class with $\alpha = 1/2$ and
$z=2$~\cite{Barabasi1995,Nicoli2009,Corwin2012,Hairer2013}.

Alternatively, the solution $u(x,t)$ can also be written in terms of its Fourier representation
\begin{equation}\label{Fourier}
u(x,t) = \sum_{k\in\ZZ}\hat{u}_k(t) e^{ikx},
\end{equation}
where $\hat{u}_k(t)$ are the Fourier components. By making use of Parseval's
identity, we can compute the expected value of $r(t)^2$ as follows:
\begin{equation}\label{SR_coef}
\left<r(t)^2\right>  = \sum_{k\in\ZZ}\left<\left|\hat{u}_k(t)\right|^2\right>-\left<\left|u_0(t)\right|^2 \right> =: \sum_{k\in\ZZ} S(k,t) - \left<\left|u_0(t)\right|^2 \right>,
\end{equation}
where we have defined the power spectral density $S(k,t) = \left<\left|\hat{u}_k(t)\right|^2 \right>$.
Therefore, if we can control the Fourier coefficients of the solution $u$, we
can control the surface roughness to evolve to a desired target value $r_d$,
i.e. $\mathrm{lim}_{t\to \infty}\sqrt{\left<r(t)^2\right>}  =  r_d$. In the
following, we propose a control methodology precisely for this purpose.

\section{Linear feedback control methodology}

The methodology we propose to control the roughness of the sKS solution
consists of two main steps. First,using a standard trick from the theory of semilinear parabolic SPDEs,
see e.g.~\cite{Ferrario2008}, we define $w$ to be the solution of the linear sKS
equation:
\begin{equation}\label{linearSKS}
w_t = -\nu w_{xxxx} - w_{xx} + \sigma\xi(x,t),
\end{equation}
and write the full solution $u$ of Eq.~(\ref{sKS}) as $u = w+v$, so that $v$ satisfies
\begin{equation}\label{deterministic}
v_t = -\nu v_{xxxx} - v_{xx} - vv_x - (vw)_x - ww_x.
\end{equation}
The important point here is to note that the above equation
\eqref{deterministic} is now a deterministic PDE with random coefficients and so
we are in a position where we can apply the methodology for nonlinear
deterministic PDEs we have developed in previous
works~\cite{IMA_paper,PRE_paper}, to stabilize its zero solution - something
possible as long as $w$ and its first derivative are bounded in an
appropriate sense (see Section~\ref{sec:Periodic proof} below for a
justification of this point). We therefore introduce the controlled equation
for $v$:
\begin{equation}
\label{deterministic_control}
v_t = -\nu v_{xxxx} - v_{xx} - vv_x - (vw)_x - ww_x + \sum_{n=-l_1}^{l_1} b^{det}_n(x)f^{det}_n(t),
\end{equation}
where $m_1 = 1+2l_1$ (with $l_1=\left[1/\sqrt{\nu}\right]$)  is the number of controls, and $b_n^{det}(x)$
are the control actuator functions. Here we use $\left[x\right]$  to denote the integer part of $x$.

Once the zero solution of the equation for $v$ has been stabilized, the second step is to control the roughness
of the solution by applying appropriate controls to the linear SPDE~(\ref{linearSKS})  for $w$ so that the solution is
driven towards the desired surface roughness $r_d$. In the following, we apply this methodology to the sKS equation,
Eq.~(\ref{sKS}) or (\ref{sKS KPZ}), by choosing two different types of controls, namely periodic controls, when the controls
are applied throughout the whole domain  and point actuated ones, when the control force is applied in a finite number
of positions in the domain.

\section{Periodic controls}\label{sec:Periodic}

\subsection{Derivation of the controlled equation}

From Eq.~(\ref{deterministic_control}),  we write
\begin{equation}\label{Fourierv}
v(x,t) = \sum_{k\in\ZZ}\hat{v}_k(t)\,e^{ikx},
\end{equation}
and take the inner product with the basis functions $e^{ikx}$ to obtain
\begin{equation}\label{controlledODEsystemv}
\dot{\hat{v}}_k =  \left(-\nu k^4 + k^2\right) \hat{v}_k + g_k(v,w) + \sum_{n=-l_1}^{l_1} b^{det}_{nk}f^{det}_n(t),
\end{equation}
with $ k\in\ZZ $ and a dot denoting a time derivative. We have introduced $b_{nk}^{det} = \int_0^{2\pi}b_n(x)e^{ikx}dx$,
and note that $g_k$ are functions of the coefficients of $v$ and $w$.

Next we define the following vectors and matrices. We denote the vector $z^v = [z_{s-} \ z_{un}^v \ z_{s+}^v ]^T$,
where $z_{un}^v = \left[v_{-l_1} \ \cdots \ v_0 \ \cdots \ v_{l_1}\right]^T$ are the coefficients of the (slow) unstable modes,
and $z_{s-}^v = \left[\cdots \ v_{-l_1-1}\right]^T$ and $z_{s+}^v = \left[v_{l_1+1} \ \cdots\right]^T$ are the coefficients of the (fast) stable modes.
We also take $G =  \left[\cdots  \ g_k \ \cdots\right]^T$,
$F^{det} = \left[f^{det}_{-l_1}(t) \ \cdots \ f^{det}_{l_1}(t)\right]^T$,
\[ A = \left[\begin{array}{ccc} A_{s-} & 0 & 0 \\ 0 & A_u & 0 \\ 0 & 0 & A_{s+} \end{array}\right] \quad \textrm{ and } \quad B^{det} = \left[ \begin{array}{c} B^{det}_{s-} \\ B^{det}_u \\  B^{det}_{s+}\end{array}\right], \]
where
\begin{align*}
& A_{s-} = \operatorname{diag}(\cdots, -(l_1+1)^4\nu +(l_1+1)^2,), \\
& A_u = \operatorname{diag}(0,-(-l_1)^4\nu +(- l_1)^2, \cdots, -l_1^4\nu + l_1^2), \\
& A_{s+} = \operatorname{diag}(-(l_1+1)^4\nu +(l_1+1)^2,\cdots),
\end{align*}
and
\begin{equation*}
B^{det}_{s-} = \left[\begin{array}{ccc}
\vdots & \cdots & \vdots\\
b_{-l_1(-l_1-2)}^{det} & \cdots & b_{l_1(-l_1-2)}^{det,s} \\
b_{-l_1(-l_1-1)}^{det} & \cdots & b_{l_1(-l_1-1)}^{det,c}
\end{array}\right],
\]
\[
B^{det}_u = \left[\begin{array}{ccc}
b_{-l_1-l_1}^{det} &\cdots & b_{l_1-l_1}^{det} \\
\vdots & \cdots & \vdots\\
b_{l_1l_1}^{det} & \cdots & b_{l_1l_1}^{det}
\end{array}\right],
\quad B^{det}_{s+} = \left[\begin{array}{ccc}
b_{-l_1(l_1+1)}^{det} & \cdots & b_{l_1(l_1+1)}^{det} \\
b_{-l_1(l_1+2)}^{det} & \cdots & b_{l_1(l_1+1)}^{det}\\
\vdots & \cdots & \vdots
\end{array}\right].
\end{equation*}
With these definitions we rewrite the infinite system of ODEs \eqref{controlledODEsystemv} as
\begin{equation}
\label{eq:v compact}
 \dot{z}^v = Az^v  + G + B^{det}F^{det}.
\end{equation}
The key point now is to note that if there exists a matrix $K^{det}$ such that all the eigenvalues
of the matrix $A_u + B^{det}_uK^{det}$ have negative real part, then the controls given by
\begin{equation}
f^{det}_n(t) = K^{det}_n z_{un}^v = K_n^{det}(z_{un}^u-z_{un}^w),
\end{equation}
where $K^{det}_{n}$ is the $n-$th row of $K^{det}$, stabilize the zero solution of
Eq.~\eqref{deterministic_control} (see \cite{IMA_paper,PRE_paper} for previously derived methodologies
for deterministic systems). The proof of this follows the same type of Lyapunov argument as for the
deterministic KS equation and is justified as long as we have nice bounds on $w$, something we will
demonstrate below.
It should be emphasized that in Eq.~(\ref{eq:v compact}) for $v$ we have
suppressed the influence of the nonlinearity on the SPDE without
assuming knowledge of its value at all times and without
changing the fundamental dynamics, in contrast to previous work~\cite{Hu2008a,Hu2008b}.

The next step is to control the stochastic linear equation for $w$ such that  the value of the second moment evolves towards a desired target.
To this end we  write
\begin{equation}\label{Fourierw}
w(x,t) = \sum_{k\in\ZZ}\hat{w}_k(t)e^{ikx},
\end{equation}
and take the inner product with the basis functions to obtain the following infinite system of ODEs for the Fourier coefficients
\begin{equation}\label{ODEsystem}
\begin{array}{rcl}
\dot{\hat{w}}_0 &=& \xi_0, \\
\dot{\hat{w}}_k & = & (-\nu k^4 + k^2) \hat{w}_k + \xi_k.
\end{array}
\end{equation}
Here $k\in\ZZ-\left\{0\right\}$, $\xi_0 = \int_0^{2\pi} \xi(x,t) \,dx$, and $\xi_k = \int_0^{2\pi} \xi(x,t)e^{ikx}\,dx$.
The solution to system \eqref{ODEsystem} is
\begin{equation}\label{solution}
\begin{array}{rcl}
\hat{w}_0(t) &=& \hat{w}_0(0) + \int_0^{t} \xi_0(t) \ dt, \\
\hat{w}_k(t) &=& e^{(-\nu k^4 + k^2)t}\hat{w}_k(0) + \int_0^{t} e^{(-\nu k^4 + k^2)(t-s)}\xi_k(s) \ ds,
\end{array}
\end{equation}
and it easily follows that
\begin{equation}\label{2ndmoment}
\left<\hat{w}_k(t)^2\right> = -\frac{\sigma^2}{2(-\nu k^4 + k^2)}(1 - e^{-2(\nu k^4 - k^2)t}), \quad k\in\ZZ.
\end{equation}
We observe that in this case the expected surface roughness only depends on the eigenvalues of the linear operator
$\mathcal{L} = -\nu \partial_x^4 - \partial_x^2$; these can be controlled using
feedback control to direct the evolution towards the desired value of surface roughness $r_d$.
Hence we introduce the controlled equation for $w$,
\begin{equation}\label{stochastic_control}
w_t = -\nu w_{xxxx} - w_{xx} + \sum_{n=-l_2, n\neq 0}^{l_2} b^{rand}_n(x)f^{rand}_n(t) +\sigma\xi(x,t),
\end{equation}
where $m_2 = 2l_2$ is the number of controls ($l_2$ needs to be larger than or equal to the number of unstable
modes and will be specified later), and we choose the functions $b_n^{rand}(x) = e^{inx}$.
We also notice that we do not need to control the eigenvalue corresponding to the constant eigenfunction ($k=0$),
since it does not contribute to the surface roughness.

By truncating the system into $N$ modes (with $N$ sufficiently large so that the contribution from higher modes can be neglected)
and taking inner products with the basis functions, we arrive  at
\begin{align}\label{controlledODEsystem}
\begin{array}{rclr}
\dot{\hat{w}}_0 &=& \xi_0,  \\
\dot{\hat{w}}_k & = & (-\nu k^4 + k^2) \hat{w}_k + f_k^{rand} + \xi_k, \quad\  k = -l_2,\dots,l_2,\\
\dot{\hat{w}}_k & = & (-\nu k^4 + k^2) \hat{w}_k + \xi_k, \quad\  k = -\frac{N}{2}, \dots,-l_2-1,l_2+1,\dots,\frac{N}{2}.
\end{array}
\end{align}
\begin{remark}
An important point to note is that  because of the choice of periodic functions for $b_n^{rand}$, the system (\ref{controlledODEsystem}) is decoupled. In fact, with such a choice of actuator functions, the matrix $B^{rand}_u$ is the identity matrix, and $B_{s\pm}^{rand}$ are zero matrices. As will be shown in Section \ref{sec:p act}, this is not the case for point actuated controls.
\end{remark}
The surface roughness for $m_2 = 2l_2$ controls is therefore given by
\begin{equation*}
\left<r^2(t)\right> =\sum_{k=-N/2, k \neq 0}^{N/2}\left<\hat{u}_k^2 (t)\right>  = \sum_{k=-l_2, k \neq 0}^{l_2}\left<\hat{u}_k^2(t) \right>  + \sum_{k=-N/2}^{-l_2-1}\left<\hat{u}_k^2(t)\right> + \sum_{k=l_2+1}^{N/2}\left<\hat{u}_k^2(t)\right>.
\end{equation*}
If we denote the desired surface roughness as $r_d^2 =\mathrm{lim}_{t\to\infty}\left<r^2(t)\right>$, we obtain
\begin{equation*}\begin{array}{rl}
r_d^2 &=\sum_{k=-l_2, k \neq 0}^{l_2}-\frac{\sigma^2|q_k|^2}{2\lambda_k}  + \sum_{k=-N/2}^{-l_2-1} -\frac{\sigma^2|q_k|^2}{2(-\nu k^4 + k^2)}+ \sum_{k=l_2+1}^{N/2} -\frac{\sigma^2|q_k|^2}{2(-\nu k^4 + k^2)} \\
& = -\frac{\sigma^2}{2}\sum_{k=-l_2, k \neq 0}^{l_2}\frac{|q_k|^2}{\lambda_k} + \underbrace{\sigma^2\sum_{k=l_2+1}^{N/2}- \frac{|q_k|^2}{-\nu k^4 + k^2}}_{\left<r_f^2\right>},
\end{array}\end{equation*}
where we have used the fact that the coefficients $q_k$ are real with $q_{-k} = q_k$ (see equation \eqref{eq:noise_Fourier}).
The chosen eigenvalues for the controlled modes are $\lambda_k$, and we take them to be $\lambda_k =\lambda$ for all $k$ to arrive at
\begin{equation}\label{e-val}
\lambda = -\frac{\sigma^2\sum_{k=1}^{l_2} |q_k|^2 }{\left<r_d^2\right> - \left<r_f^2\right>}.
\end{equation}
To control the surface roughness we therefore define the controls $f^{rand}_k$ such that the new eigenvalues satisfy the following relation
\begin{equation}\label{controls_formula}
f^{rand}_k = \left(\lambda +\nu k^4-k^2\right)\hat{w}_k.
\end{equation}
Finally, putting Eqs.~\eqref{deterministic_control} and \eqref{stochastic_control} together, yields the controlled equation
for the full solution $u$
\begin{equation}\label{controlledsKS}
u_t = -\nu u_{xxxx} - u_{xx} - uu_x + \xi(x,t) + \sum_{n=-l_1}^{l_1}b^{det}_n(x)f_n^{det}(t) +  \sum_{n=-l_2}^{l_2}b^{rand}_n(x)f_n^{rand}(t).
\end{equation}

\subsection{Proof of applicability of the control methodology}
\label{sec:Periodic proof} Our aim here is to prove that the solution $v$ can
indeed be controlled to zero even though Eq.~(\ref{deterministic_control})
has random coefficients, i.e. the terms $(vw)_x$ and $w w_x$. We will show that by
adopting a similar argument as used for the proof of existence and uniqueness
of solutions of the sKS equation \cite{Ferrario2008}, we can apply a
Lyapunov-type argument as in the deterministic KS equation \cite{PRE_paper}.

We use \eqref{controls_formula} to write the solution of Eq.~\eqref{stochastic_control} as
\[
w(t) = e^{\mathcal{A}t}w(x,0) + \sigma\int_0^t e^{\mathcal{A}(t-s)}d\xi(s),
\]
where $\mathcal{A} = -(\nu A^2-A + F)$, $A = -\partial_x^2$ and $F$ is an operator discretised as
\[
F =  \left[\begin{array}{ccc} 0 & 0 & 0 \\ 0 & \operatorname{diag}(\lambda - \nu k^4 + k^2) & 0 \\ 0 & 0 & 0\end{array}\right].
\]
We take $\mathcal{G}$ to be a trace class operator, so that it satisfies~\cite[Assumption (3.1)]{Ferrario2008}.
Writing
\[
e^{\mathcal{A}(t-s)}\xi(s) = \sigma\sum_{j,k\in\ZZ} q_k \,e^{-(\nu k^4 - k^2 + f_k)(t-s)}<e_k,e_j>\beta_k(s)e_j,
\]
we have
\begin{subequations}
\begin{align}
\label{1stMoment} \mathbb{E}[w(t)] &= e^{\mathcal{A}t}w(x,0) = 0,\\
\mathbb{E}[|w(t)-\mathbb{E}[w(t)]|^2] &= \sigma^2\sum_{j,k\in\ZZ}\int_0^t e^{-2(\nu k^4-k^2 + f_k)(t-s)}|q_k|^2\,|<e_k,e_j>|^2, \nonumber \\
\label{2ndMoment} & = \sum_{k=-l_2}^{l_2} \frac{\sigma^2|q_k|^2}{\lambda} + \sum_{|k|\geq l_2} \frac{\sigma^2|q_k|^2}{2(\nu k^4 - k^2)} = r_d^2,
\end{align}
\end{subequations}
where we used $<e_k,e_j> = 0$ and $f_k = \lambda + \nu k^4 - k^2$. Since we are assuming that the covariance
matrix $G$ is such that assumption (3.1) in \cite{Ferrario2008} is satisfied, we have that $w(t) \in \dot{L}^2(0,2\pi)$,
the space of mean zero $L^2$ functions, almost surely, for any time $t$. This also means~\cite{Prato2014} that there
exists a continuous version of $w$ that we shall consider from now on.

Now we define $B(u,v) = uv_x$ and $b(u,v,w) = <B(u,v),w> = \int_0^{2\pi}uv_xw \ dx $,
which satisfy the following relations \cite{Ferrario2008,Robinson2001}:
\begin{subequations}\label{b_relations}
\begin{align}
\label{b1} \|b(u_1,u_2,u_3)\|_{L^2} & \leq \|u_1\|_{L^2}\|u_{2,x}\|_{L^\infty}\|u_3\|_{L^2} \leq c\|u_1\|_{L^2}\|Au_2\|_{L^2}\|u_3\|_{L^2},  \\
\label{b2} b(u,u,u) & = 0,    \\
\label{b3} b(u_1,u_2,u_2) & = b(u_2,u_2,u_1)  = -\frac{1}{2}b(u_2,u_1,u_2),  \\
\label{b4} b(u_1,u_2,u_3) & = - b(u_2,u_1,u_3) - b(u_1,u_3,u_2).
\end{align}
\end{subequations}
and  \cite[Proposition (2.1)]{Ferrario2008}:
\begin{subequations}\label{B_relations}
\begin{align}
\label{B1} \|B(u,v)\|_{D(A^{-1})} &\leq c\|Au\|_{L^2}\|v\|_{L^2},\\
\label{B2} \|B(z,v)\|_{D(A^{-1})} &\leq c\|u\|_{L^2}\|Av\|_{L^2},\\
\label{B3}  \|B(z,z)\|_{D(A^{-1})} &\leq c\|z\|_{L^2}^2,\\
\label{B4} \|B(u,v)\|_{D(A^{-\delta})} &\leq c\|u\|_{D(A^{\frac{1}{2}-\delta})}\|v\|_{D(A^{\frac{1}{2}-\delta})}.
\end{align}
\end{subequations}
On the other hand, we notice that the existence of the matrix $K^{det}$ implies that the operator $\mathcal{A}$,
such that $\mathcal{A}v = -\nu v_{xxxx} - v_{xx} - \sum_{n=-l_1}^{l_1} b_n^{det}(x)f_n^{det}(t)$, satisfies
\begin{equation}\label{Lyapunov}
\int_0^{2\pi} v\mathcal{A}v \ dx \leq -a\|v\|_{L^2}^2,
\end{equation}
for some positive constant $a$, which in turn depends on the eigenvalues we choose for the controlled operator.
Therefore, multiplying equation \eqref{deterministic_control} by $v$ and integrating by parts yields
\begin{multline}
\frac{1}{2}\frac{d}{dt}\|v\|_{L^2}^2 \leq -a\|v\|_{L^2}^2 - \overbrace{b(v,v,v)}^{=0} - b(v,w,v) - b(v,w,v) - b(w,w,v)\\
= -a\|v\|_{L^2}^2 + b(w,v,v) + \frac{1}{2}b(w,v,w) \leq -a\|v\|_{L^2}^2 +c\|w\|_{L^2}\|v\|_{L^2}\|Av\|_{L^2}\\
+ \frac{c}{2}\|w\|_{L^2}^2\|Av\|_{L^2} \leq -\left(a-\frac{c}{2}\|w\|_{L^2}^2\right)\|v\|_{L^2}^2 + c\|Av\|_{L^2}^2 + \frac{c}{2}\|w\|_{L^2}^4,
\end{multline}
where we have used Young's inequality and relations \eqref{b_relations} and \eqref{B_relations}. The term $c\|Av\|_{L^2}^2$ can be controlled using sufficiently strong controls and the
last term on the right-hand-side is a constant that depends on the desired surface roughness and that again can be controlled by choosing large enough eigenvalues. Therefore, by choosing the controls such that $a$ is large enough, $\|v\|^2_{L^2}$ is a Lyapunov function for this system and the zero solution for the controlled equation for $v$ is stable.

\subsection{Numerical results}

We apply now the methodology presented above with periodic controls to the
sKS with either the Burgers nonlinearity (Eq.~(\ref{sKS})) or the KPZ
nonlinearity (Eq.~(\ref{sKS KPZ})). For simplicity, we consider white noise
in both space and time ($q_k=1$). All our numerical experiments are
solved using spectral methods in space and a second-order backward
differentiation formula  scheme in time ~\cite{Akrivis2011}.
\begin{figure}[t]
\centering
\includegraphics[width=1.0\linewidth]{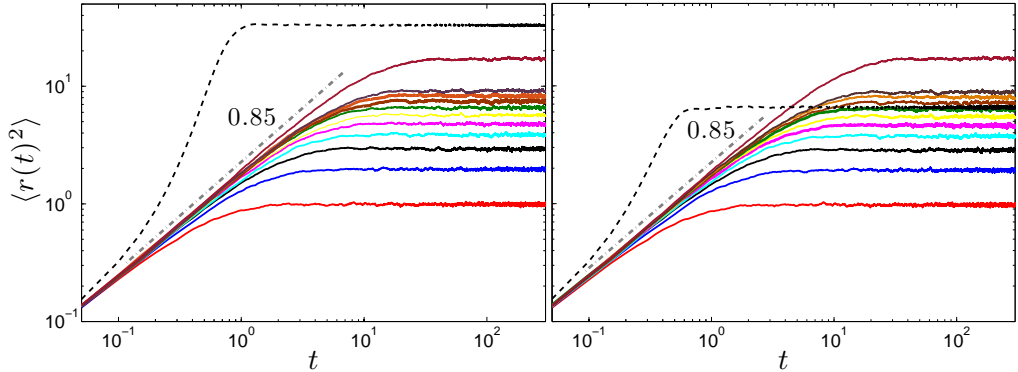}\label{fig:KS}
\caption{Squared value of the surface roughness of the solutions to the sKS equation with Burgers nonlinearity (left) and the KPZ nonlinearity (right) for $\nu = 0.05$, $\sigma = 0.5$ and different values of the desired surface roughness, ranging from $1$ to $10$, and $20$. The dashed lines show the value of the uncontrolled roughness, and the straight dashed-line corresponds to a guide-to-eye line with slope $0.85$. }
\label{fig:log-log}
\end{figure}

\subsubsection{Controlling the roughening process}
\begin{figure}[t]
\centering
\includegraphics[width=1.0\linewidth]{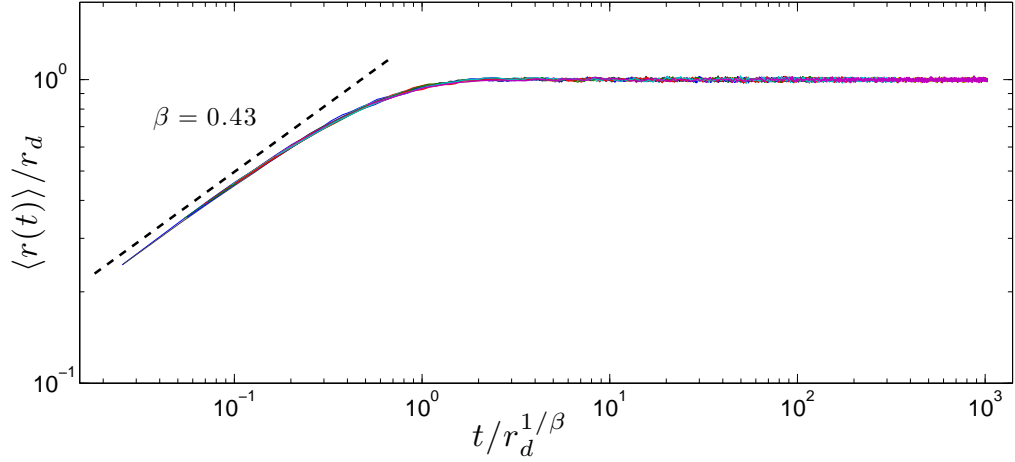}\label{fig:KS}
\caption{Surface roughness rescaled by the target value $r_d$ against the rescaled time $t/r_d^{1/\beta}$ for all cases shown in Fig.~\ref{fig:log-log}. The dashed line corresponds to a guide-to-eye line with slope $0.43$.}
\label{fig:colapse}
\end{figure}

We solved Eqs.~(\ref{sKS}) and (\ref{sKS KPZ}) for $\nu = 0.05$ and
$\sigma=0.5$, controlling its solutions towards various desired surface
roughnesses $r_d$. The results are presented in Fig.~\ref{fig:log-log}. We
observe that in both cases the solution exhibits a power-law behavior at
short times of the form given by Eq.~(\ref{Beta}) until the solution
saturates to the desired value of the roughness. It is interesting to note that
the exponent in all cases is the same with $\beta \approx 0.43$, independently of
the type of nonlinearity and desired surface roughness (note that the exponent in
Fig.\ref{fig:log-log} is $\approx 0.85 = 2\beta$, since we are plotting $<r(t)^2>$).
This becomes even
clearer if time and surface roughness are rescaled by their saturation
values, $t_s$ and $r_d$, respectively. By noting that $r_d\sim t_s^\beta$,
Eq.~(\ref{Beta}) is rewritten as:
\begin{equation}
\label{Beta scaled}
\frac{\langle r(t)\rangle}{r_d} \sim  \left\{\begin{array}{ll}
          x^\beta &  \mathrm{if}\ \ x\ll 1, \\
          1 &   \mathrm{if}\ \ x\gg 1, \end{array}\right.
\end{equation}
where $x = t/r_d^{1/\beta}$. Fig.~\ref{fig:colapse} shows that all the
different cases  presented in Fig.~\ref{fig:log-log} collapse into a single
curve which is given by \eqref{Beta scaled} above with the universal value $\beta \approx
0.43$.

\begin{figure}[t!]
 \centering
 \includegraphics[width=0.9\linewidth]{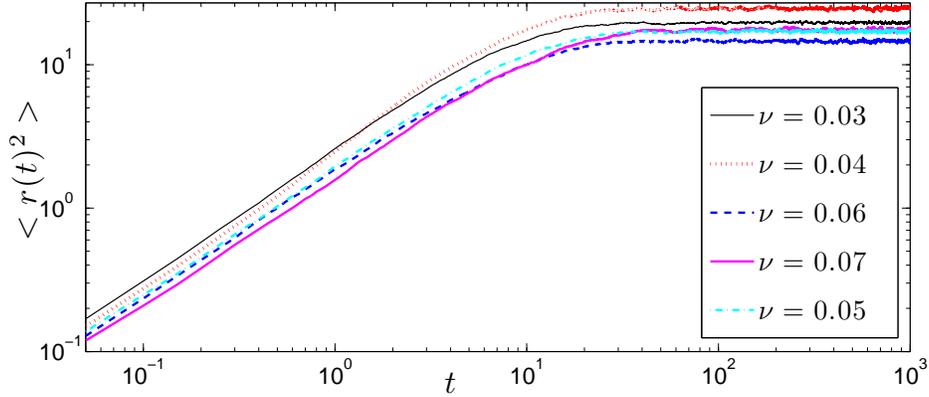}\label{fig:domain}
\caption{Controlled roughness with same target value $r_d^2 = 20$ and different values of $\nu$ - the domain size
increases as $\nu$ decreases.}
\label{fig:diff nu}
 \end{figure}

We also study the effect of changing the domain by varying the parameter
$\nu$. Fig.~\ref{fig:diff nu} shows the numerical results obtained when we
fix the target value $r_d$ and change the parameter $\nu$. We observe that
changing the domain does not change the growth rate (we observe the same
growth exponent $\beta\approx 0.43$) but it does slightly affect the final value
of the roughness.
An important point to note is that since we are controlling the surface
roughness of the solution $r(t)$ to be at a specified value $r_d$, the saturated state in which
the statistical properties become stationary, is reached whenever $r(t) =r_d$. Therefore,
the saturation time, and the long-time roughness value,  should not depend on the system size.

\begin{figure}[t!]
 \centering
 \includegraphics[width=0.9\linewidth]{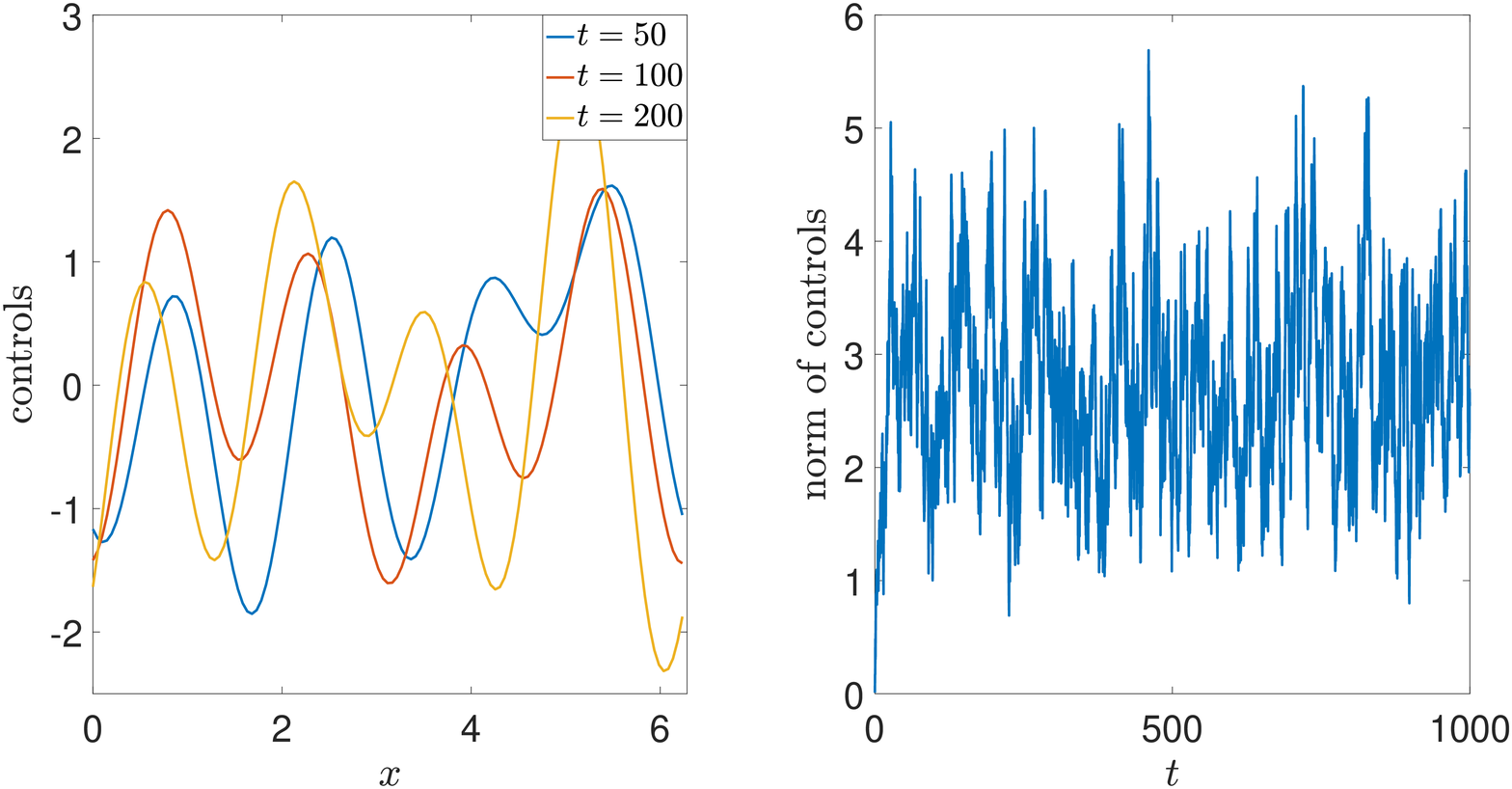}\label{fig:controls1}
\caption{Left: Controls at different time steps. Right: $L^2$-norm of the controls as a function of time. For both figures
$\nu = 0.05$ and $\sigma = 0.5$.}
 \end{figure}

\begin{figure}[t!]
 \centering
 \includegraphics[width=0.9\linewidth]{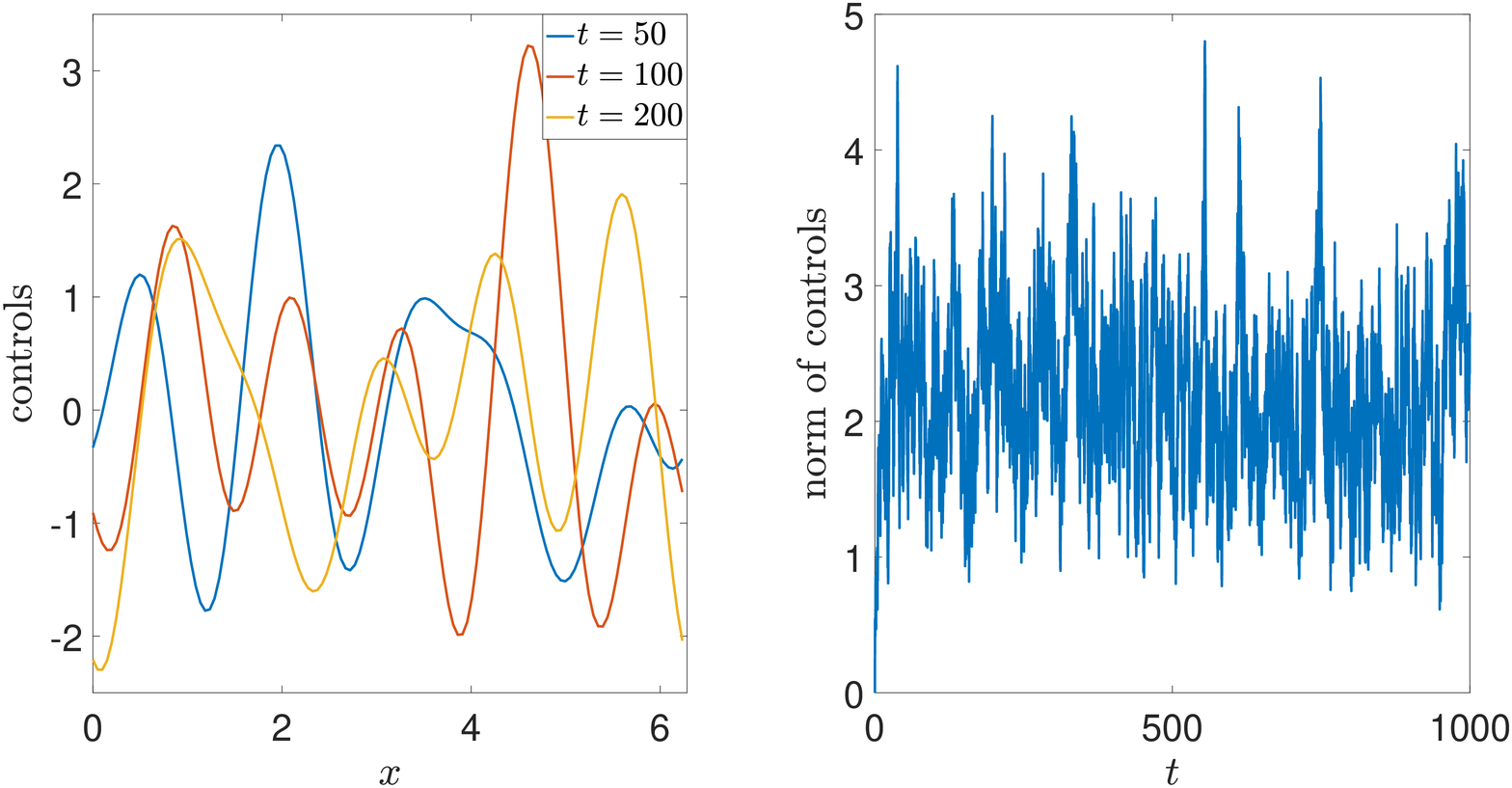}\label{fig:controls2}
\caption{Left: Controls at different time steps. Right: $L^2$-norm of the controls as a function of time. For both figures
$\nu = 0.03$ and $\sigma = 0.5$.}
 \end{figure}

\subsubsection{Changing the shape of the solution}
It is important to emphasize that in addition to controlling the roughness of the
solution of the sKS equation, we can also change its shape, something that could have
ramifications in
technological applications such as materials processing. We quantify this by considering the
surface roughness of the solution as is its distance
to the desired state. If $\bar{u}(x)$ is the ultimate desired shape of the solution,
then the quantity we are trying to control now becomes
\begin{equation}\label{controlShape}
r(t)=\sqrt{\frac{1}{2\pi}\int_0^{2\pi}\left(u(x,t) - \bar{u}\right)^2 \ dx}.
\end{equation}
Using Parseval's identity we compute the expected value of $r(t)^2$
\begin{equation}\label{SR_coef}
\left<r(t)^2\right> =\sum_{k\in\ZZ, k\neq 0}\left<\left(u_k(t)-\bar{u}_k\right)^2 \right>.
\end{equation}
\begin{figure}[t!]
\centering
\includegraphics[width=0.9\linewidth]{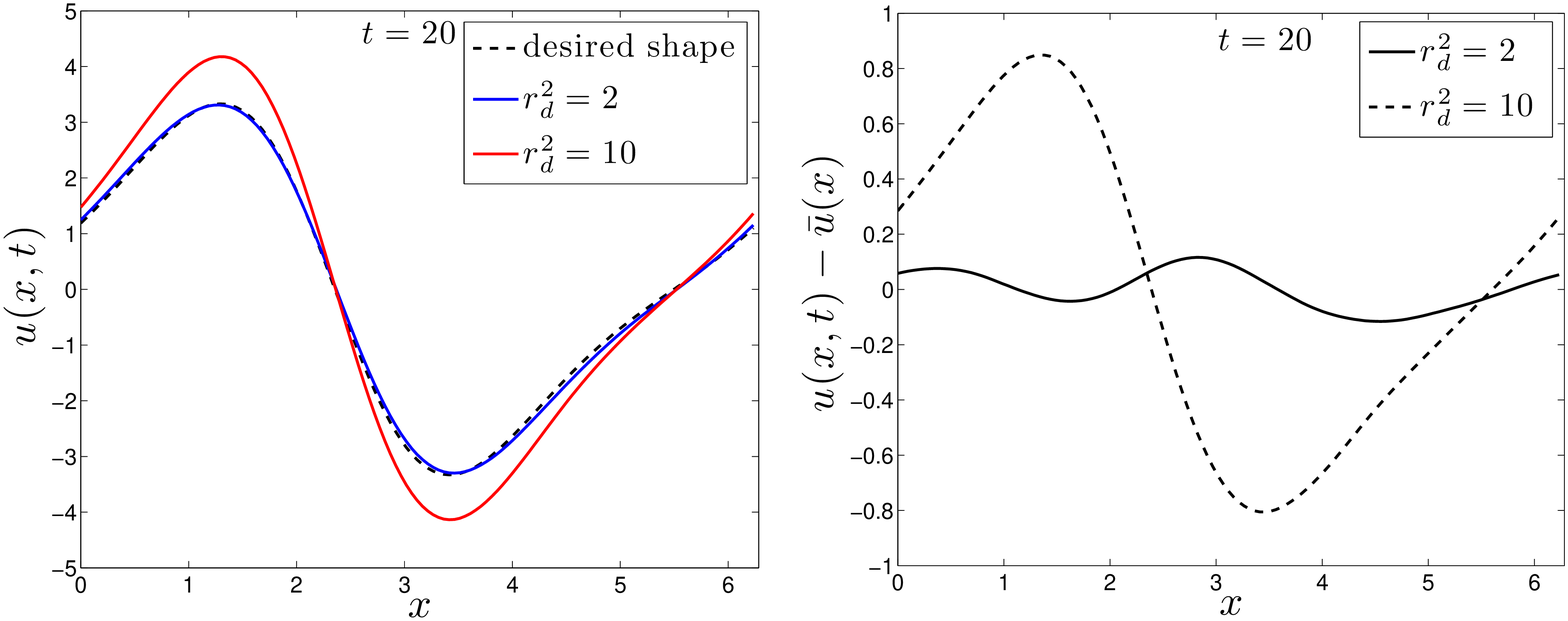}\\
\includegraphics[width=0.9\linewidth]{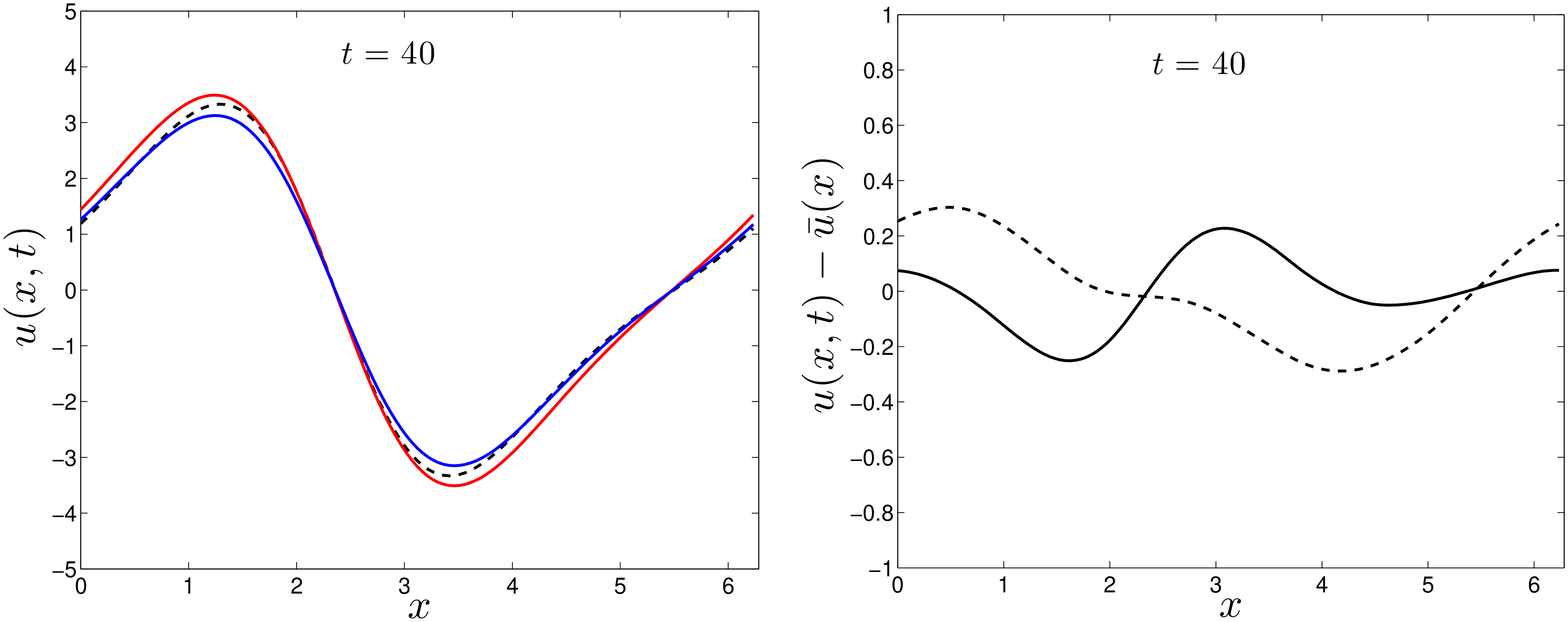}\\
\includegraphics[width=0.9\linewidth]{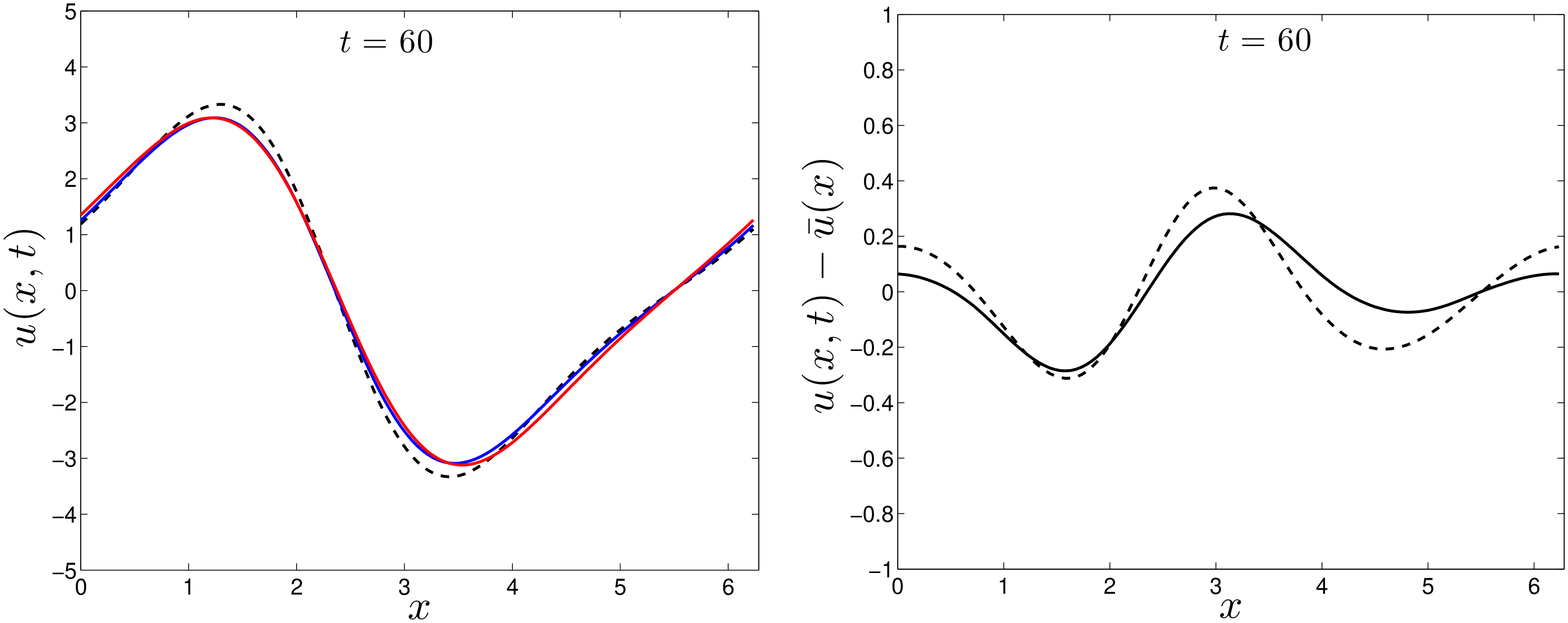}
\caption{Snapshots of the sKS equation solution  controlled to the shape of one of the steady states of the KS equation (left panels) and difference between current solution and desired shape for two different desired surface roughness (right panels). Parameters are $\nu = 0.5$, $\sigma = 0.5$, $r_d^2= 2$ (blue) and $r_d^2 = 10$ (red) with $T = 100$ and $dt = 5\times 10^{-3}$.}
\label{fig:shape}
 \end{figure}
To control the shape of the solution, we can therefore control the solution of  equation (\ref{deterministic_control}) for $v$ to the desired shape rather than controlling it to zero. This in turn implies the use of
$f^{det}_n(t) = K^{det}_n \left(z_{un}^v-z_{un}^{\bar{u}}\right) = K_n^{det}(z_{un}^u-z_{un}^w-z_{un} ^{\bar{u}})$.
We use the steady states of the KS equation for a chosen value of $\nu$ to
define the desired shape $\bar{u}$.  Results are shown in
Fig.~\ref{fig:shape} for $\nu=0.5$, where we can see that the solution is fluctuating
around the imposed shape.

\section{Point actuated controls}
\label{sec:p act}
We now consider controls that are point actuated and not distributed throughout the whole domain, i.e.~the
functions $b_n(x)$ are now given by $b_n(x)= \delta(x-x_n)$, where $\delta(x)$ is the Dirac delta function.
By repeating the same procedure as with periodic controls, writing $w = \sum_{k\in\ZZ}\hat{w}_ke^{ikz}$ and taking
the inner product with the eigenfunctions of the linear operator, $\mathcal{L} = -\nu \partial_x^4 - \partial_x^2$,
we obtain the following infinite system of linear stochastic ODEs
\begin{equation}\label{controlledODEsystemPointActuators}
\begin{array}{rclr}
\dot{\hat{w}}_0 &=& \xi_0 + \sum_{n=1}^m b_n^0 f_n, & \\
\dot{\hat{w}}_k & = & (-\nu k^4 + k^2) \hat{w}_k +  + \sum_{n=1}^{m_2} b_n^k f_n + \xi_k, & k \neq 0,
\end{array}
\end{equation}
 where the coefficients $b_n^k$ are defined from the functions $b(x) = \delta(x-x_n)$ as before,  $b_n^k = \int_0^{2\pi}b_n(x)e^{ikx}dx$.
We can  see that the difference between the above system and the periodic controls one given by \eqref{controlledODEsystem},
is that now the system is coupled. In fact the coupling matrix is not symmetric, and most importantly, it does not commute with its transpose. Therefore the solution does not follow directly and we cannot easily write the second moment of the coefficients as a function of the eigenvalues as in the previous section. To obtain the controlled equation we thus need to apply a different approach.

Let the controls $F = [f_1, \cdots, f_m]$ be such that $F = K\hat{w}$ where $\hat{w}$ is a vector containing the
Fourier coefficients of $w$, and the matrix $K$ is to be determined.
Since the equations are not decoupled we cannot multiply by $w$ and integrate to find directly the second moment of the coefficients.
However, we can make use of results derived in \cite{Jimenez2015} which provide simplified formulas for the first and second moments of  systems analogous to \eqref{controlledODEsystemPointActuators}.
Let $\Xi$ be the vector $\Xi_k = \xi_k$ and $C = A+BK$ where $A = \operatorname{diag}-\nu k^4 + k^2$
 and $B_{kn} = b_n^k$, so that we can write the truncated system \eqref{controlledODEsystem} as
\[
\dot{\hat{w}} = A\hat{w} + BK\hat{w} + \Xi := C\hat{w} + \Xi.
\]
We also assume without loss of generality that $\mathbf{m}(0) = \mathbb{E}(\hat{w}(0)) = 0$ and $\mathbf{P}(0) = \mathbb{E}
(\hat{w}(0)\hat{w}(0)^T)= 0$.  Then Theorem $4$ in \cite{Jimenez2015} states that
\[
\mathbf{m}(t) = \mathbb{E}(\hat{w}(t)) = 0 \quad \text{ and } \quad \mathbf{P}(t) = \mathbb{E}(\hat{w}(t)\hat{w}(t)^T) = \mathbf{H}_1\mathbf{F}_1^T + \mathbf{F}_1\mathbf{H}_1^T
\]
where $\mathbf{F_1}$ and $\mathbf{H_1}$ are the $(1,1)$ and $(1,3)$ blocks of the matrix $e^{Mt}$ where in the case of space-time white noise, $M$ is
\begin{equation*}
M=\left[\begin{array}{cccc}
C & 0 & \frac{\sigma^2}{2}I & 0 \\
0 & 0 & 0 & 0 \\
0 & 0 & -C^T & 0 \\
0 & 0 & 0 & 0
\end{array}
\right],
\end{equation*}
$I$ is an appropriately sized identity matrix and the zeros stand for zero matrices of appropriate size.
We compute $e^{Mt}$ and conclude that
\[
\mathbf{F}_1 = e^{Ct},
\]
and
\begin{align*}
\mathbf{H}_1 = & \frac{\sigma^2}{2}\left[It + (C-C^T)\frac{t^2}{2} + (C^2 - CC^T + (C^T)^2)\frac{t^3}{3!}\right] \\
& + \frac{\sigma^2}{2}\left[(C^3 - C^2C^T + C(C^T)^2 - (C^T)^3)\frac{t^4}{4!} + \cdots\right].
\end{align*}
Since $\mathbf{F}_1\mathbf{H}_1^T = (\mathbf{H}_1\mathbf{F}_1^T)^T$ and  $(\mathbf{H}_1\mathbf{F}_1^T)^T = \mathbf{H}_1\mathbf{F}_1^T$, we have $ \mathbf{H}_1\mathbf{F}_1^T + \mathbf{F}_1\mathbf{H}_1^T = 2 \mathbf{H}_1\mathbf{F}_1^T$, from which we obtain
\begin{multline}
\mathbf{P}(t) = \sigma^2\left(It + (C+C^T)\frac{t^2}{2} + \left(C^2 + 2 CC^T + (C^T)^2\right)\frac{t^3}{3!}+\right. \\
+ \left(C^3 + 3 C^2C^T + 3C(C^T)^2 + (C^T)^3\right)\frac{t^4}{4!} +\\
\left. +\left(C^4 + 4C^3C^T + 6C^2(C^T)^2 + 4 C(C^T)^3 + (C^T)^4\right)\frac{t^5}{5!} + \cdots\right).
\end{multline}

\begin{remark}
In the periodic case, the matrix $C$ is diagonal, so $CC^T = C^TC$, and this is exactly the same as
\begin{equation}
\mathbf{P}(t) = \sigma^2\sum_{n=1}^\infty \left(C+C^T \right)^{n-1}\frac{t^n}{n!}.
\end{equation}
In addition, when choosing the eigenvalues of $C$, we can ensure that it is invertible and therefore $C+C^T$ is also invertible, which gives
\begin{equation}
\mathbf{P}(t) = -\sigma^2\left(C+C^T\right)^{-1} + \sigma^2 e^{(C+C^T)t},
\end{equation}
so as $t\rightarrow\infty$, $\mathbf{P}(t)\rightarrow - \sigma^2\left(C+C^T\right)^{-1}$ and
\begin{equation}
<r(t)^2> = \operatorname{tr}(\mathbf{P}(t))\rightarrow \sum_{k\in{\ZZ-\left\{0\right\}}}-\frac{\sigma^2}{2\lambda_k},
\end{equation}
where $\lambda_k$ are the chosen eigenvalues of $C$. Hence we recover the same result as before.
\end{remark}
It is important to note that the matrix $C$ is not normal, i.e.~it does not
commute with its transpose, and the eigenvalues of $C+C^T$ do not satisfy the
useful properties that allow us to obtain (30). However, we are not interested in
knowing the full matrix $\mathbf{P}(t)$, but only its trace
\begin{multline}
\operatorname{tr}(\mathbf{P}(t)) = \operatorname{tr}\left(\sigma^2\left(It + (C+C^T)\frac{t^2}{2} + \left(C^2 + 2 CC^T + (C^T)^2\right)\frac{t^3}{3!}+\right. \right.\\
+ \left(C^3 + 3 C^2C^T + 3C(C^T)^2 + (C^T)^3\right)\frac{t^4}{4!} +\\
\left.\left. +\left(C^4 + 4C^3C^T + 6C^2(C^T)^2 + 4 C(C^T)^3 + (C^T)^4\right)\frac{t^5}{5!} + \cdots\right)\right).
\end{multline}
By now making use of  the linearity of the trace and its continuity to pass it inside the infinite sum, we get
\begin{multline}
\operatorname{tr}(\mathbf{P}(t)) =\sigma^2\left( \operatorname{tr}(I)t +  \operatorname{tr}(C+C^T)\frac{t^2}{2} +  \operatorname{tr}\left(C^2 + 2 CC^T + (C^T)^2\right)\frac{t^3}{3!}+ \right.\\
+  \operatorname{tr}\left(C^3 + 3 C^2C^T + 3C(C^T)^2 + (C^T)^3\right)\frac{t^4}{4!} +\\
\left. + \operatorname{tr}\left(C^4 + 4C^3C^T + 6C^2(C^T)^2 + 4 C(C^T)^3 + (C^T)^4\right)\frac{t^5}{5!} + \cdots\right).
\end{multline}
We also note that
\[
\operatorname{tr}\left(C^2 + 2 CC^T + (C^T)^2\right) = \operatorname{tr}\left(C^2 +  CC^T + C^TC+ (C^T)^2\right) = \operatorname{tr}(C+C^T)^2.
\]
Similarly we can prove that the terms multiplied by $\frac{t^n}{n!}$ are of the form
$\operatorname{tr}\left((C+C^T)^{n-1}\right)$ and we finally obtain
\begin{multline}
\operatorname{tr}(\mathbf{P}(t)) =\sigma^2\left( \operatorname{tr}(I)t +  \operatorname{tr}(C+C^T)\frac{t^2}{2} +  \operatorname{tr}\left(C+C^T\right)^2\frac{t^3}{3!}+ \right.\\
\left.+  \operatorname{tr}\left(C+C^T\right)^3\frac{t^4}{4!} + \operatorname{tr}\left(C+C^T\right)^4\frac{t^5}{5!} + \cdots\right).
\end{multline}
We proceed by assuming that $C+C^T$ is invertible, so that we can multiply by $I =(C+C^T)^{-1} (C+C^T)$ and add and subtract pertinent terms to obtain
\begin{equation}\label{trace_eqn}
\operatorname{tr}(\mathbf{P}(t)) = -\sigma^2\operatorname{tr}\left(C+C^T\right)^{-1} + \sigma^2\operatorname{tr}\left(\left(C+C^T\right)^{-1}\sum_{n\in\NN} \left(C+C^T\right)^n\frac{t^n}{n!} \right).
\end{equation}

\begin{remark}
This does not change the proof provided in Section \ref{sec:Periodic proof}, it
only changes the formula for the covariance so that the bounds are still
valid.
\end{remark}

\begin{remark} We emphasize that the following assumptions were made here:
\begin{itemize}
\item[(a)] $C+C^T$ needs to be invertible.
\item[(b)] In order for the surface roughness to converge to a finite value, we require all of the eigenvalues of $C+C^T$ to be negative,
so that the exponential part disappears.
\end{itemize}
\end{remark}

\subsection{Computation of the matrix $K$}
We note in Eq.~\eqref{trace_eqn} that we now need to control the trace of $D^{-1} =
(C+C^T)^{-1}$ and we can do that by prescribing the eigenvalues of $D$.
Hence we can control the surface roughness by finding a matrix $K$ such
that the eigenvalues of
\begin{equation}\label{point_control}
D = C+C^T = A+BK + A^T +  (BK)^T = 2A + BK + K^TB^T,
\end{equation}
are a given set $\left\{\mu_1,\dots,\mu_N\right\}$.
Since we only wish to prescribe the eigenvalues of $D$, rather than knowing all of its entries, we can tackle this problem by using the information provided by the characteristic polynomial, $\chi_D$, of $D$.
We know  that
\begin{equation}\label{eq:char_1}
\chi_D(t) = \prod_{i=1}^N (t-\mu_i) = \sum_{k=0}^N(-1)^k\sum_{J: |J| = k}\prod_{j\in J} \mu_j t^{N-k},
\end{equation}
where $J$ is a subset of $\left\{1,\dots,N\right\}$. Equivalently we can express $\chi_D$ in terms of the sum over all its diagonal minors, i.e.,
\begin{equation}\label{eq:char_2}
\chi_{D} = \sum_{k=0}^N (-1)^k\,\eta_k\, t^{N-k},
\end{equation}
where $\eta_k$ is the sum over all of the diagonal minors of size $k$ of $D$. This translates into a system of $N$ nonlinear algebraic equations,
\[
\eta_k = \sum_{J: |J| = k}\prod_{j\in J} \mu_j,
\]
for the $m\times N$ entries of the matrix $K$ - see
\cite{Stephen_algorithm} for details on the solution to this problem. For the
purposes of our study, we will make use a nonlinear solver (e.g.,
\textsc{Matlab}'s \emph{fsolve}) to obtain the matrix $K$. Given the
structure of the matrix $B$ and the fact that the system is underdetermined,
convergence is rather slow when solving the problem directly. We overcome
this by performing a change of variables: we obtain the SVD decomposition of
$B$ by finding matrices $X$ and $Y$ such that $\tilde{B} = XBY^T$, and
multiply equation \eqref{point_control} by $X^T$ on the left and by $X$ on
the right. We then define $\tilde{K} = Y^TKX$, $\tilde{A} = X^TAX$ and
$\tilde{D} = X^TDX$, so that we obtain the equation
\begin{equation}\label{point_control_new_basis}
\tilde{D} = 2\tilde{A} + \tilde{B}\tilde{K} + \tilde{K}^T\tilde{B}^T.
\end{equation}
This is of the same form as \eqref{point_control}, but where the matrix $\tilde{B}$ is diagonal.
We find that this accelerates the convergence of the system (for moderate values of $N$)
and we were able to get satisfactory numerical results, which we now present.

\begin{figure}[t!]
 \centering
 \includegraphics[width=1\linewidth]{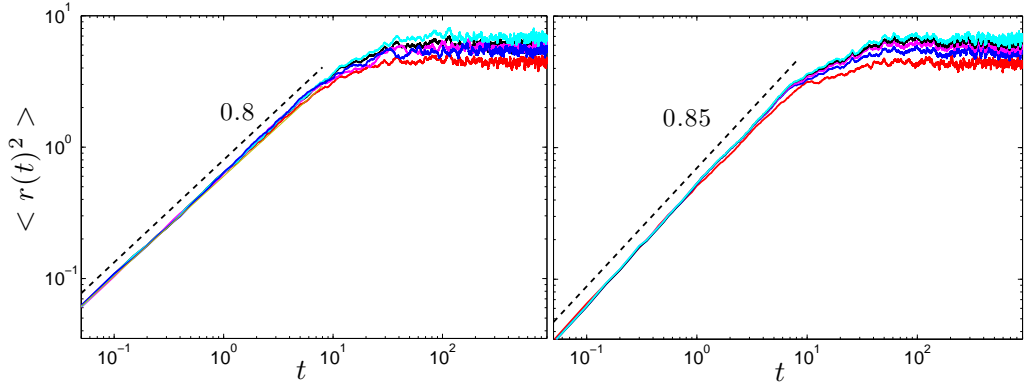}
\caption{Squared value of the surface roughness of the solutions to the sKS equation with Burgers nonlinearity for $\nu = 0.04$, $\sigma = 0.5$ and different values of the desired surface roughness, ranging from $2$ to $6$. Left: using space time white noise; Right: using colored noise described by the coefficients $q_k = |k|^{-1}$. We applied $m=3$ point actuated controls, which were located at the positions $x_1 = \frac{\pi}{3}, \, x_2 = \pi, \, x_3 = \frac{5\pi}{3}$.}
\label{fig:point_actuated}
 \end{figure}

\subsection{Numerical results}

We apply the methodology presented in the previous section with point actuated controls to the sKS with the Burgers
nonlinearity [Eq.~(\ref{sKS})] (similar results are expected for the KPZ nonlinearity [Eq.~(\ref{sKS KPZ})]).
We solved Eq.~\eqref{sKS} for $\nu = 0.4$ and $\sigma = 0.5$. For this value
of $\nu$ the linear operator has $3$ unstable modes and we apply $m=3$
controls. We note that even though we do not need to control the mode
corresponding to the first moment of the solution when using periodic
controls, we benefit from doing so in this case, since the matrix $D$ would
not be invertible if we allowed for a zero eigenvalue. We consider  either
space-time white noise ($q_k = 1$) or colored noise with $q_k = |k|^{-1}$,
(which is chosen to decay at a fast rate so that the system can be truncated at a smaller value of N)
and control the solution towards various desired values $r_d$ of the surface
roughness.

The results are depicted in Fig.~\ref{fig:point_actuated} where we observe
that the solution still exhibits a power-law behavior with similar exponent
as in the periodic case (there we found $\beta\approx 0.43$) until it saturates at the
desired value of the surface roughness. We note that even though we obtained
satisfactory results for the range of values of $r_d^2$ selected in Fig.~\ref{fig:point_actuated},
further increase of $r_d$ does not lead to the expected saturated results.
This may be due to the
relatively small system truncation value $N=21$ that was found necessary in order to
obtain convergence of the problem to find the entries of the matrix $K$. Further work is
required in this direction that is beyond the scope of the present study.

\section{Conclusions}

We have presented a generic methodology for controlling the surface roughness of nonlinear SPDEs exemplified by the sKS
equation with either the Burgers nonlinearity or the KPZ nonlinearity and using periodic or point actuated controls.

We have shown that with the appropriate choice of periodic controls the
solutions of these equations can be forced to have a wide range of prescribed
surface roughness values, defined to be the distance of the solution to its mean
value. We are also able to force the solutions to a prescribed shape given
by steady state solutions of the deterministic KS equation. We find that the
solution to the controlled problem exhibits a power-law behavior with a
universal exponent $\beta\approx0.43$ , which
is not affected by changes in the length of the domain, and is found to be independent of the type of nonlinearity of the sKS equation.

When using point actuated feedback controls, the problem becomes considerably
harder to solve due to the fact that the resulting system of linear ODEs is
not decoupled. This leads to the need to solve a new matrix problem which is
similar to a matrix Lyapunov equation; to the best of our knowledge such a problem
has not been tackled before. The complexity of this problem makes it harder
to solve for a large system truncation value $N$, but we have obtained satisfactory results when
controlling towards a range of surface roughness values for moderate $N$. The
study of this matrix problem is an interesting separate problem and our
detailed results and associated algorithm for its solution can be found in
\cite{Stephen_algorithm}.

We believe that our framework offers several distinct advantages over
other approaches. First,
the controls we derived are linear functions of the solution $u$, and this in
turn decreases the computational cost of their determination. Second,
our splitting methodology allows us to deal with the nonlinear term
directly rather than including it in the controls, thus rendering the resulting
equation essentially linear and easier to handle.

One interesting observation is that feedback control methodologies can be
used, in principle, in order to accelerate the convergence of infinite
dimensional stochastic systems such as the sKS and the KPZ equations to their
steady state. This might prove to be a useful computational tool when
analyzing the equilibrium properties of such systems, e.g. calculating
critical exponents, studying their universality class etc. Accelerating
convergence to equilibrium and reducing variance by adding appropriate
controls that modify the dynamics while preserving the equilibrium states has
already been explored for Langevin-type samplers that are used in molecular
dynamics~\cite{LelievreNierPavliotis2013, DuncanLelievrePavliotis2016}.
In addition, it would be interesting to investigate how our
methodology could be used to control the kinetic roughening process of the
system. In particular, our results show that the dynamics  towards saturation
is described  in terms of power laws. Whether we can control the values of
the associated scaling exponents during such scale-free behaviour is
something that requires a systematic study of different stochastic models by
controlling them to evolve towards large values of the surface roughness. We
shall examine these and related issues for the sKS and the KPZ equations in
future studies.

\section*{Acknowledgments}
We are grateful to the anonymous reviewer for insightful comments
and valuable suggestions. We acknowledge financial support from Imperial
College through a Roth Ph.D. studentship, the Engineering and Physical
Sciences Research Council of the UK through Grants No. EP/H034587, No.
EP/J009636, No. EP/K041134, No. EP/L020564, No. EP/L024926, No. EP/L025159,
No. EP/L027186, and No. EP/K008595 and the European Research Council via
Advanced Grant No. 247031.

\bibliographystyle{plain}
 \bibliography{biblio}

\begin{thebibliography}{10}

\bibitem{Akrivis2011}
G.~Akrivis, D.~T. Papageorgiou, and Y.-S. Smyrlis.
\newblock Linearly implicit methods for a semilinear parabolic system arising
  in two-phase flows.
\newblock {\em IMA Journal of Numerical Analysis}, 31:299--321, 2011.

\bibitem{Alava2004}
M.~Alava, M.~Dub\'{e}, and M.~Rost.
\newblock Imbibition in disordemaroon media.
\newblock {\em Adv. Phys.}, 53(2):83--175, 2004.

\bibitem{Barabasi1995}
A.-L. Barabasi and H.~E. Stanley.
\newblock {\em Fractal {C}oncepts in {S}urface {G}rowth}.
\newblock Cambridge University Press, 1995.

\bibitem{Bloemker2002}
D.~Bl\"{o}mker, C.~Gugg, and M.~Raible.
\newblock Thin-film growth models: roughness and correlation functions.
\newblock {\em Eur. J. Appl. Math.}, 13:385--402, 2002.

\bibitem{Bouchbinder2006}
Eran Bouchbinder, Itamar Procaccia, St\'ephane Santucci, and Lo\"{\i}c Vanel.
\newblock Fracture surfaces as multiscaling graphs.
\newblock {\em Phys. Rev. Lett.}, 96:055509, Feb 2006.

\bibitem{Buceta1997}
J.~Buceta, J.~Pastor, M.~A. Rubio, and F.~J. de~la Rubia.
\newblock The stochastic {K}uramoto-{S}ivashinsky equation: a model for compact
  electrodeposition growth.
\newblock {\em Phys. Lett. A}, 235:464--468, 1997.

\bibitem{Buceta1998}
J.~Buceta, J.~Pastor, M.~A. Rubio, and F.~J. de~la Rubia.
\newblock Small scale properties of the stochastic stabilized
  {K}uramoto-{S}ivashinsky equation.
\newblock {\em Physica D}, 113:166--171, 1998.

\bibitem{Corwin2012}
I.~Corwin.
\newblock The {K}ardar-{P}arisi-{Z}hang equation and universality class.
\newblock {\em Random Matrices: Theory and Appl.}, 1:1130001, 2012.

\bibitem{Cuerno1995}
A.~Cuerno, H.~A. Makse, S.~Tomassone, S.~T. Harrington, and H.~E. Stanley.
\newblock Stochastic erosion for surface erosion via ion sputtering: Dynamical
  evolution from ripple morphology to rough morphology.
\newblock {\em Phys. Rev. Lett.}, 75:4464--4467, 1995.

\bibitem{Cuerno1995a}
R.~Cuerno and A.-L. Barabasi.
\newblock Dynamic scaling of ion-sputtemaroon surfaces.
\newblock {\em Phys. Rev. Lett.}, 74(23):4746--4749, 1995.

\bibitem{Diez2016}
J.~A. Diez and A.~G. Gonz\'{a}lez.
\newblock Metallic-thin-film instability with spatially correlated thermal
  noise.
\newblock {\em Phys. Rev. E}, 93:013120, 2016.

\bibitem{Duan2001}
J.~Duan and V.~J. Ervin.
\newblock On the stochastic {K}uramoto-{S}ivashinsky equation.
\newblock {\em Nonlinear Anal.-Theor.}, 44:205--216, 2001.

\bibitem{DuncanLelievrePavliotis2016}
A.~B. Duncan, T.~Lelievre, and G.~A. Pavliotis.
\newblock Variance maroonuction using nonreversible langevin samplers.
\newblock {\em J. Stat. Phys.}, 2016.

\bibitem{Elezgaray1994}
J.~Elezgaray, G.~Berkooz, and P.~Holmes.
\newblock Large-scale statistics of the {K}uramoto-{S}ivashinsky equation: A
  wavelet-based approach.
\newblock {\em Phys. Rev. E}, 54:224--230, Jul 1996.

\bibitem{Ferrario2008}
B.~Ferrario.
\newblock Invariant measures for a stochastic {K}uramoto-{S}ivashinsky
  equation.
\newblock {\em Stoch. Anal. Appl.}, 26(2):379--407, 2008.

\bibitem{IMA_paper}
S.~N. Gomes, D.~T. Papageorgiou, and G.~A. Pavliotis.
\newblock Stabilising nontrivial solutions of the generalised
  {K}uramoto-{S}ivashinsky equation using feedback and optimal control.
\newblock {\em IMA J. Appl. Math. doi: 10.1093/imamat/hxw011}, 2016.

\bibitem{PRE_paper}
S.~N. Gomes, M.~Pradas, S.~Kalliadasis, D.~T. Papageorgiou, and G.~A.
  Pavliotis.
\newblock Controlling spatiotemporal chaos in active dissipative-dispersive
  nonlinear systems.
\newblock {\em Phys. Rev. E}, 92:022912, 2015.

\bibitem{Stephen_algorithm}
S.~N. Gomes and S.~J. Tate.
\newblock On the solution of a {L}yapunov type matrix equation arising in the
  control of stochastic partial differential equations.
\newblock {\em Submitted to IMA J. Appl. Math.}, 2016.

\bibitem{Gruen2006}
G.~Gr\"{u}n, K.~Mecke, and M.~Rauscher.
\newblock Thin-film flow influenced by thermal noise.
\newblock {\em J. Stat. Phys.}, 122(6):1261--1294, 2006.

\bibitem{Hairer2013}
M.~Hairer.
\newblock Solving the {KPZ} equation.
\newblock {\em Annals of Maths}, 178(2):559--664, 2013.

\bibitem{Harrison2016}
M.P. Harrison and R.M. Bradley.
\newblock {P}roducing virtually defect-free nanoscale ripples by ion
  bombardment of rocked solid surfaces.
\newblock {\em Phys. Rev. E}, 93:040802(R), 2016.

\bibitem{Hu2008a}
G.~Hu, Orkoulas G, and P.~D. Christofides.
\newblock Stochastic modeling and simultaneous regulation of surface roughness
  and porosity in thin film deposition.
\newblock {\em Ind. Eng. Chem. Res.}, 48:6690--6700, 2009.

\bibitem{Hu2008b}
G.~Hu, Y.~Lou, and P.~D. Christofides.
\newblock Dynamic output feedback covariance control of stochastic dissipative
  partial differential equations.
\newblock {\em Chem. Eng. Sci.}, 63:4531--4542, 2008.

\bibitem{Hu2009}
G.~Hu, G.~Orkoulas, and P.~D. Christofides.
\newblock Modeling and control of film porosity in thin film deposition.
\newblock {\em Chem. Eng. Sci.}, 64:3668--3682, 2009.

\bibitem{Hu2009a}
G.~Hu, G.~Orkoulas, and P.~D. Christofides.
\newblock Regulation of film thickness, surface roughness and porosity in thin
  film growth using deposition rate.
\newblock {\em Chem. Eng. Sci.}, 64:3903--3913, 2009.

\bibitem{Jimenez2015}
J.C. Jimenez.
\newblock Simplified formulas for the mean and variance of linear stochastic
  differential equations.
\newblock {\em Applied Mathematics Letters}, (49):12--19, 2015.

\bibitem{Kalliadasis2012}
S.~Kalliadasis, C.~Ruyer-Quil, B.~Scheid, and M.~G. Velarde.
\newblock {\em Falling Liquid Films}, volume 176.
\newblock Springer, 2012.

\bibitem{Kardar1986}
M.~Kardar, G.~Parisi, and Y.-C.Zhang.
\newblock Dynamic scaling of growing interfaces.
\newblock {\em Phys. Rev. Lett.}, 56:889--892, Mar 1986.

\bibitem{Krug1997}
J.~Krug.
\newblock Origins of scale invariance in growth processes.
\newblock {\em Adv. Phys.}, 46(2):139--282, 1997.

\bibitem{Lauritsen1996}
K.~B. Lauritsen, R.~Cuerno, and H.~A. Makse.
\newblock Noisy {K}uramoto-{S}ivashinsky equation for an erosion model.
\newblock {\em Phys. Rev. E}, 54(4):3577--3580, 1996.

\bibitem{LelievreNierPavliotis2013}
T.~Lelievre, F.~Nier, and G.~A. Pavliotis.
\newblock Optimal non-reversible linear drift for the convergence to
  equilibrium of a diffusion.
\newblock {\em J. Stat. Phys.}, 152(2):237--274, 2013.

\bibitem{Lou2005}
Y.~Lou and P.~D. Christofides.
\newblock Feedback control of surface roughness in sputtering processes using
  the stochastic {K}uramoto-{S}ivashinsky equation.
\newblock {\em Comput. Chem. Eng.}, 29:741--759, 2005.

\bibitem{Lou2006}
Y.~Lou and P.~D. Christofides.
\newblock Nonlinear feedback control of surface roughness using a stochastic
  {PDE}: {D}esign and application to a sputtering process.
\newblock {\em Industrial \& Engineering Chemistry Research}, 45:7177--7189,
  2006.

\bibitem{Lou2008}
Y.~Lou, G.~Hu, and P.~D. Christofides.
\newblock Model pmaroonictive control of nonlinear stochastic partial
  differential equations with application to a sputtering process.
\newblock {\em Process Systems Engineering}, 54(8):2065--2081, 2008.

\bibitem{Lou2009}
Y.~Lou, G.~Hu, and P.~D. Christofides.
\newblock Model pmaroonictive control of nonlinear stochastic {PDE}s:
  Application to a sputtering process.
\newblock {\em American Control Conference}, 2009.

\bibitem{Nesic2015}
S.~Nesic, R.~Cuerno, E.~Moro, and L.~Kondic.
\newblock {F}ully nonlinear dynamics of stochastic thin-film dewetting.
\newblock {\em Phys. Rev. E}, 92:061002(R), 2015.

\bibitem{Nicoli2009}
M.~Nicoli, R.~Cuerno, and M.~Castro.
\newblock Unstable nonlocal interface dynamics.
\newblock {\em Phys. Rev. Lett.}, 102:256102, Jun 2009.

\bibitem{Pradas2006}
M.~Pradas and A.~Hern\'andez-Machado.
\newblock Intrinsic versus superrough anomalous scaling in spontaneous
  imbibition.
\newblock {\em Phys. Rev. E}, 74:041608, Oct 2006.

\bibitem{Pradas2012}
M.~Pradas, G.~A. Pavliotis, S.~Kalliadasis, D.~T. Papageorgiou, and
  D.~Tseluiko.
\newblock Additive noise effects in active nonlinear spatially extended
  systems.
\newblock {\em Eur. J. of Appl. Math.}, 23:563--591, 2012.

\bibitem{Pradas2011}
M.~Pradas, D.~Tseluiko, S.~Kalliadasis, D.~T. Papageorgiou, and G.~A.
  Pavliotis.
\newblock Noise induced state transitions, intermittency, ans universality in
  the noisy {K}uramoto-{S}ivashinsky equation.
\newblock {\em Physical Review Letters}, 106:060602--, 2011.

\bibitem{Prato2014}
G.~Da Prato and J.~Zabczyk.
\newblock {\em Stochastic equations in infinite dimensions}.
\newblock Cambridge University Press, second edition, 2014.

\bibitem{Procaccia1992}
I.~Procaccia, M.~H. Jensen, V.~S. L'vov., K.~Sneppen, and R.~Zeitak.
\newblock Surface roughening and the long-wavelength properties of the
  {K}uramoto-{S}ivashinsky equation.
\newblock {\em Phys. Rev. A}, 46:3220--3224, Sep 1992.

\bibitem{Robinson2001}
J.~C. Robinson.
\newblock {\em Infinite-Dimensional Dynamical Systems. An introduction to
  dissipative parabolic PDEs and the theory of global attractors}.
\newblock Cambridge University Press, 2001.

\bibitem{Rost1995}
M.~Rost and J.~Krug.
\newblock Anisotropic {K}uramoto-{S}ivashinsky equation for surface growth and
  erosion.
\newblock {\em Phys. Rev. Lett.}, 75(21):3894--3897, 1995.

\bibitem{Soriano2005}
J.~Soriano, A.~Mercier, R.~Planet, A.~Hern\'andez-Machado, M.~A.
  Rodr\'{\i}guez, and J.~Ort\'{\i}n.
\newblock Anomalous roughening of viscous fluid fronts in spontaneous
  imbibition.
\newblock {\em Phys. Rev. Lett.}, 95:104501, Aug 2005.

\bibitem{Thompson2016}
A.~B. Thompson, S.~N. Gomes, G.~A. Pavliotis, and D.~T. Papageorgiou.
\newblock Stabilising falling liquid film flows using feedback control.
\newblock {\em Phys. Fluids}, 28:012107, 2016.

\bibitem{Yakhot1981}
V.~Yakhot.
\newblock Large-scale properties of unstable systems governed by the
  {K}uramoto-{S}ivashinksi equation.
\newblock {\em Phys. Rev. A}, 24:642--644, Jul 1981.

\bibitem{Zhang2010}
X.~Zhang, G.~Hu, G.~Orkoulas, and P.~D. Christofides.
\newblock Pmaroonictive control of surface mean slope and roughness in a thin
  film deposition process.
\newblock {\em Chem. Eng. Sci.}, 65:4720--4731, 2010.

\end{thebibliography}

\end{document}